\documentclass[doublecol,final]{hex}

\usepackage{amsfonts,amssymb,amsmath,graphicx}

\title{Structures in 3D double-diffusive convection  and
possible approach to the Saturn's polar hexagon modeling}
\shorttitle{Structures in 3D double-diffusive convection} %Insert here a short version of the title if it exceeds 70 characters

\author{S. B. Kozitskiy}
\shortauthor{S. B. Kozitskiy}

\institute{Il'ichev Pacific Oceanological Institute, Baltiyskaya 43, Vladivostok, 690041, Russia}


\abstract{ Three-dimensional double-diffusive convection in a
horizontally infinite layer of an uncompressible fluid interacting
with horizontal vorticity field is considered in the neighborhood of
Hopf bifurcation points. A family of amplitude equations for
variations of convective cells amplitude is derived by
multiple-scaled method. Shape of the cells is given as a
superposition of a finite number of convective rolls with different
wave vectors. For numerical simulation of the obtained systems of
amplitude equations a few numerical schemes based on modern ETD
(exponential time differencing) pseudo-spectral methods were
developed. The software packages were written for simulation of
roll-type convection and convection with square and hexagonal type
cells. Numerical simulation has showed that the convection takes the
form of elongated ``clouds'', ``spots'' or ``filaments''. It was
noted that in the system quite rapidly a state of diffusive chaos is
developed, where the initial symmetric state is destroyed and the
convection becomes irregular both in space and time. The obtained
results may be the basis for the construction of more advanced
models of multi-component convection, for instance, model of
Saturn's polar hexagon.}

\newcommand{\Div}{\mathop{\rm div}\nolimits}
\newcommand{\ex}{\mathop{\rm e}\nolimits}
\newcommand{\exi}[1]{\mathop{\rm e}\nolimits^{{\mathrm i}#1}}
\newcommand{\sico}[1]{\left\{\begin{array}{c}{\sin{#1}}\\{\cos{#1}}\end{array}\right\}}
\newcommand{\diag}{\mathop{\rm diag}\nolimits}
\newcommand{\cj}{\mathop{\rm c.c.}\nolimits}
\newcommand{\irm}{\mathrm i}
\newcommand{\Ns}{\widehat{N}}
\newcommand{\Ls}{\widehat{L}}
\newcommand{\Ms}{\widehat{M}}
\newcommand{\Ds}{\mathcal{D}}
\begin{document}

\maketitle

\section{Introduction}

It is believed that the convection is the most common case of gas
and liquid flows in the Universe~\cite{Getling}. Among the various
types of convection the so called {\it double-diffusive} convection
holds a special place. Physical systems with double-diffusive
convection have two components with significantly different
coefficients of diffusion. It can be heat and salt in the sea water,
heat and helium in stellar atmospheres, or two reagents in chemical
reactors. As a result of various spatial distribution of these
components in a gravitational field the convection arises, which can
have various forms and lead to a variety of
phenomena~\cite{Huppert,Radko}.
In oceanography thermohaline convection plays an important role in
heat and mass transfer processes in the ocean and affect different
small-scale processes that lead to the formation of vertical fine
structure~\cite{KABC}.

During last 50 years double-diffusive convection is actively studied
by both experimental and theoretical methods, including numerical
modeling.
One of the classical methods to study the system with convective
instability near the bifurcation points is the method of amplitude
equations. For the case of Rayleigh-Benard convection this method
was used by Newell and Whitehead~\cite{Newell}. It allowed to reduce
the original PDE system to a nonlinear evolution equation for one
roll mode. Also it made possible to obtain the equations for the
case of several roll modes with nonlinear interaction, so that the
shape of the convective cells can be an arbitrary. Since then, the
method of amplitude equations is frequently used to study various
convective phenomena.

In 80-90 years the formation of structures in the neighborhood of
Hopf bifurcation points for the horizontally translation-invariant
systems was actively studied in some works. The development of
oscillations in such systems gives rise to different types of waves
(eg, standing, running, modulated, chaotic), which is well described
by a complex Ginzburg-Landau equations (CGLE).
The equations of this type must be derived from the basic system of partial
differential equations for the given physical system by asymptotic
methods. However, a full and well-grounded derivation of amplitude
equations for systems with double-diffusive convection (especially
three-dimensional) is still poorly represented in the literature.

For the 2D roll-type double-diffusive convection the amplitude
equations of CGLE type were firstly derived and studied numerically
in the work~\cite{Bretherton}. The amplitude equations for the case
of roll-type Rayleigh-Benard convection were derived in the
work~\cite{Zippelius}.

The main idea of the present article consists in combining strict
mathematical derivation of amplitude equations by multiple-scaled
method (following \cite{K10}) and considering arbitrary number of
interacting roll-type convective modes over horizontal vorticity
field (as it was done for the Rayleigh-Benard convection
in~\cite{Newell,Zippelius}) for obtaining the amplitude equations
for three-dimensional double-diffusive system in the neighborhood of
Hopf bifurcation points. This also develops the ideas of previous
work~\cite{KABC,K11,K15,K16}, where a two-dimensional and
three-dimensional convection with a square-type and roll-type cells
was investigated by alike methods. Then the derived amplitude
equations are investigated numerically. Possible forms of 3D
double-diffusive convection are described.

\section{Formulation of the problem and basic equations}

Consider 3D double-diffusive convection in a liquid layer of a width
$h$, confined by two plane horizontal boundaries. The liquid layer is
heated and salted from below. The governing equations in this case
are hydrodynamical equations for a liquid mixture in the
gravitational field~\cite{Lan}:
\begin{equation*}\label{meq}
\begin{split}
& \partial_t{\bf v} + ({\bf v}\nabla){\bf v}=
-\rho^{-1}\nabla p + \nu\Delta{\bf v}+{\bf g}\,,\\
& \partial_t T + ({\bf v}\nabla) T = \chi\Delta T\,,\\
& \partial_t S + ({\bf v}\nabla) S = D\Delta S\,,\\
& \Div{\bf v} = 0\,.
\end{split}
\end{equation*}
Where ${\bf v}(t,x,y,z)$ is the velocity field of liquid, $T(t,x,y,z)$
is the temperature, $S(t,x,y,z)$ is the salt concentration, $p(t,x,y,z)$
is the pressure, $\rho(t,x,y,z)$ is the density of
liquid, ${\bf g}$ is the acceleration of gravity, $\nu$ is the kinematic
viscosity of fluid, $\chi$ is the thermal diffusivity of the liquid, $D$ is
the salt diffusivity. Cartesian frame with the horizontal $x$-axis and
$y$-axis is used, while the $z$-axis is directed upward and $t$ is the
time variable.

Distributed sources of heat and salt are absent. On the upper and
lower boundaries of the layer the constant values of temperature and
salinity are supported, the higher ones are at the lower boundary.

The governing equations are transformed into dimensionless form with
the use of Boussinesq approximation and following units for length,
time, velocity, pressure, temperature and salinity are respectively:
$h$, $h^2/\chi$, $\chi/h$, $\rho_0\chi^2/h^2$, $T_{\Delta}$,
$S_{\Delta}$, where $T_{\Delta}$ and $S_{\Delta}$ are temperature
and salinity differences across the layer.
The dimensionless governing equations for momentum and diffusion of
temperature and salt are~\cite{K11}:
\begin{equation}\label{meq3}
\begin{split}
& u_t+(uu_x+vu_y+wu_z)=-p_x+\sigma\Delta u\,,\\
& v_t+(uv_x+vv_y+wv_z)=-p_y+\sigma\Delta v\,,\\
& w_t+(uw_x+vw_y+ww_z)=-p_z+\sigma\Delta w\\
& \qquad\qquad+\sigma R_T\theta-\sigma R_S\xi\,,\\
& \theta_t+(u\theta_x+v\theta_y+w\theta_z)-w=\Delta\theta\,,\\
& \xi_t+(u\xi_x+v\xi_y+w\xi_z)-w=\tau\Delta\xi\,,\\
& u_x + v_y + w_z = 0\,.
\end{split}
\end{equation}
Where $\sigma=\nu_0/\chi$ is the Prandtl number ($\sigma\approx7.0$),
$\tau=D/\chi$ is the Lewis number ($0<\tau< 1$,
usually $\tau=0.01-0.1$).
$R_T = ({g{\alpha'}h^3}/\chi\nu)T_{\Delta}$
and
$R_S = ({{g}{\gamma'}{h^{3}}}/{\chi\nu})S_{\Delta}$
are the temperature and the salinity Rayleigh numbers,
$\alpha'$ and $\gamma'$ are cubic expansion coefficients.
Fluid velocity field is represented by the vector
${\bf v}(t,x,y,z)=(u,v,w)^T$ with superscript <<T>> denoting
transposition.
Variables $\theta(t,x,y,z)$ and $\xi(t,x,y,z)$ denote deviations
of temperature and salinity from their stationary linear profiles, so
\begin{equation*}
\begin{split}
& T(t,x,y,z)=T_{+}+T_{\Delta}[\theta(t,x,y,z)-z]\,,\\
& S(t,x,y,z)=S_{+}+S_{\Delta}[\xi(t,x,y,z)-z]\,.
\end{split}
\end{equation*}
$T_{+}$ and $S_{+}$ are the temperature and salinity at the lover boundary
of the area.

Free-slip boundary conditions are used for the dependent variables
(the horizontal velocity component is undefined):
$$
u_z = v_z = w = \theta = \xi = 0\quad\text{at}\quad z=0,\,1\,.
$$
It is believed that they are suitable to describe the convection in
the inner layers of liquid and do not change significantly the
convective instability occurrence criteria for the investigated
class of systems~\cite{Weiss}.

\section{Derivation of amplitude equations - general frame of decomposition}

Con\-si\-der the equations for double-diffusive convection in the
vicinity of a bifurcation point, the temperature and salinity
Rayleigh numbers for which are designated as $R_{Tc}$ and $R_{Sc}$ respectively.
In this case the Rayleigh numbers can be represented as follows:
$$R_T=R_{Tc}(1+\varepsilon^2 r_T), \qquad R_S=R_{Sc}(1+\varepsilon^2 r_S)\,.$$
At least one of the values $r_T$ or $r_S$ is of unit order, and the small
parameter $\varepsilon$ shows how far from the bifurcation point the system is.
In the case when the system is destabilized by increasing the temperature
gradient in the layer we have $r_T=1$ and $r_S=0$.
Respectively $R_T=R_{Tc}(1+\varepsilon^2)$ and $R_S=R_{Sc}$. According to
these expressions the small parameter can be defined by formula:
$$\varepsilon=\sqrt{\frac{R_T-R_{Tc}}{R_{Tc}}}\,.$$

To derive the amplitude equations we use the derivative-expansion
method~\cite{Nayfeh}, which is the case of the multiple-scale method.
Introduce the slow variables:
$$
T_1 = \varepsilon t\,,\quad T_2=\varepsilon^2 t\,,\quad X =
\varepsilon x\,,\quad Y = \varepsilon y\,.
$$
In accordance with the chosen method we assume that the dependent
variables now depend on $t$, $T_1$, $T_2$, $x$, $y$, $z$, $X$, which
are considered as independent. Also we replace the derivatives in the
equations (\ref{meq3}) for the prolonged ones by the rules:
$$
\partial_t \rightarrow \partial_t + \varepsilon\partial_{T_1}
 + {\varepsilon}^{2}\partial_{T_2}\,, \quad
\partial_x \rightarrow \partial_x + \varepsilon\partial_{X}\,,\quad
\partial_y \rightarrow \partial_y + \varepsilon\partial_Y\,.
$$
Then the equations (\ref{meq3}) can be written as:
\begin{equation}\label{meq4}
\Ls\varphi=-\varepsilon\Ls_1\varphi-\varepsilon^2\Ls_2\varphi -
\Ns_1(\varphi,\varphi)-\varepsilon\Ns_2(\varphi,\varphi)\,.
\end{equation}
Where we have introduced vector of the dependent variables
$\varphi = (u,v,w,\theta,\xi,p)^T$ and matrix-differential operators $\Ls$\,,
$\Ls_1$ and $\Ls_2$:
\begin{equation*}
\begin{split}
& \Ls = L_{a}\partial_t-L_{b}\Delta_{\perp}+L_{c}\partial_x+L_{d}\partial_y+L_{e1}\partial_z\\
& \qquad\qquad-L_{g}-\sigma R_{Tc}L_{R1}+\sigma R_{Sc}L_{R2}\,,\\
& \Ls_1 =
L_{a}\partial_{T_1}-2L_{b}(\partial_x\partial_X+\partial_y\partial_Y)
+L_{c}\partial_X+L_{d}\partial_Y\,, \\
& \Ls_2 = L_{a}\partial_{T_2}-L_{b}\Delta_{\perp} -\sigma r_T
R_{Tc}L_{R1}+\sigma r_S R_{Sc}L_{R2}\,.
\end{split}
\end{equation*}
Here $\Delta_{\perp}=\partial_X^2+\partial_Y^2$.
Matrices $L_i(6\times6)$ have the following nonzero elements:
\begin{equation*}
\begin{split}
& L_a = \diag(1,1,1,1,1,0)\,,\qquad L_b = \diag(\sigma,\sigma,\sigma,1,\tau,0)\,,\\
& L_c(1,6)=1\,,\quad L_c(6,1)=1\,,\qquad L_d(2,6)=1\,,\\
& L_d(6,2)=1\,,\quad L_{e1}(3,6)=1\,,\quad L_{e1}(6,3)=1\,,\\
& L_{e2}(3,6)=-1\,,\quad L_{e2}(6,3)=1\,,\quad L_g(4,3)=1\,,\\
& L_g(5,3)=1\,,\quad L_{R1}(3,4)=1\,,\quad L_{R2}(3,5)=1\,.
\end{split}
\end{equation*}
Also nonlinear operators $\Ns_1$ and $\Ns_2$ are introduced as the following vectors:
\begin{equation*}
\begin{split}
& \Ns_k(\varphi_i,\varphi_j) = (\Ms_k(\varphi_i,u_j),\Ms_k(\varphi_i,v_j),\Ms_k(\varphi_i,w_j),\\
& \qquad\qquad\qquad\qquad\qquad\qquad \Ms_k(\varphi_i,\theta_j),\Ms_k(\varphi_i,\xi_j),0)^T\,,\\
& \Ms_1(\varphi_i,u_j)= u_iu_{jx}+v_iu_{jy}+w_iu_{jz}\,,\\
& \Ms_2(\varphi_i,u_j)= u_iu_{jX}+v_iu_{jY}\,.
\end{split}
\end{equation*}

We seek solutions of equations (\ref{meq4}) in the form of asymptotic
series in powers of small parameter $\varepsilon$:
\begin{equation}\label{sets}
\varphi = \sum_{i=1}^{\infty}\varepsilon^i\varphi_i
=\varepsilon\varphi_1+\varepsilon^2\varphi_2+\varepsilon^3\varphi_3+\cdots\,.
\end{equation}
After their substitution in (\ref{meq4}) and collection the terms at
$\varepsilon^n$ we obtain the systems of equations to determine the
terms of the series (\ref{sets}).
\begin{equation}\label{sys13}
\begin{split}
%----------------------------------------------------------------
& O(\varepsilon):\hphantom{^2}\quad \Ls\varphi_1 = 0\,,  \\
%----------------------------------------------------------------
& O(\varepsilon^2):\quad \Ls\varphi_2 =
- \Ls_1\varphi_1-\Ns_1(\varphi_1,\varphi_1)\,, \\
%----------------------------------------------------------------
& O(\varepsilon^3):\quad \Ls\varphi_3 =
- \Ls_1\varphi_2-\Ls_2\varphi_1
-\Ns_1(\varphi_1,\varphi_2)\\
& \qquad\qquad\qquad\qquad -\Ns_1(\varphi_2,\varphi_1)-\Ns_2(\varphi_1,\varphi_1)\,.
%----------------------------------------------------------------
\end{split}
\end{equation}
In addition to systems (\ref {sys13}) at the powers of $\varepsilon$
from the first to third in some cases it makes sense to consider the systems
at higher powers of the small parameter, for example, to include in the
final amplitude equations a nonlinear terms of the fifth order.

However, in this article we restrict ourselves to the equations
obtained at no higher than $\varepsilon^3$.
Thus linear equations at $\varepsilon^1$ will give us the form of solution
as the sum of normal modes and conditions for the absence of secular terms
in the systems at $\varepsilon^2$ and $\varepsilon^3$ will lead to equations
on the amplitudes of each of the normal modes.

\section{The terms of the first order in $\varepsilon$}

At $O(\varepsilon^1)$  we obtain the following system:
\begin{equation}\label{seq1}
\begin{split}
\Ls\varphi_1 = 0\,.
\end{split}
\end{equation}
This linear system has a solution in the form of sum of $n$ normal modes (convective rolls):
\begin{equation}\label{nmode1a}
\begin{split}
&\varphi_1= \sum_{j=1}^n{\varphi}_{1j} + \widehat{\varphi}_1 + \cj\\
& \qquad = \sum_{j=1}^n A_{j}(X,Y,T_1,T_2)\check{\varphi}_{1j}
\ex^{\lambda t}\exi{\vec{k}_j\cdot\vec{x}}\sico{\pi z}\\
& \qquad\qquad\qquad\qquad + \widehat{\varphi}_1(X,Y,T_1,T_2) + \cj\,.
\end{split}
\end{equation}
The cosine in the braces is selected for variables $u_1,v_1,p_1$,
in another cases the sine is selected. Vectors $\vec{k}_j$ have components
$\vec{k}_j=(k_{aj},k_{bj})$.
Components of the vectors $\check{\varphi}_{1j}$ and $\widehat{\varphi}_1$
are:
\begin{equation*}
\begin{split}
& \check{\varphi}_{1j} = \\
& \left(\frac{\irm k_{aj}\pi}{k^2}\,,\frac{\irm k_{bj}\pi}{k^2}\,,1\,,
\frac{1}{\lambda+\varkappa^2}\,,\frac{1}{\lambda+\tau\varkappa^2}\,,
-\frac{\pi}{k^2}(\lambda+\sigma\varkappa^2) \right)\,,\\
& \widehat{\varphi}_1 = (\widehat{u}_1\,,\widehat{v}_1\,,0\,,0\,,0\,,\widehat{p}_1)\,.
\end{split}
\end{equation*}
Without the great loss of generality we omit $\widehat{w}_1$,
$\widehat{\theta}_1$ and $\widehat{\xi}_1$, which as the other
members with caps have sense of integration constants on slow
horizontal variables. More detailed analysis shows that these
terms are zero or do not lead to a physically meaningful results.
The terms $\widehat{u}_1$ and $\widehat{v}_1$ form the velocity field,
against which the convection develops.

Components of
$\check{\varphi}_{1j}$ are obtained by substitution of the anzats
(\ref{nmode1a}) into equations (\ref{seq1}), and it is true
$L_j\check{\varphi}_{1j}=0$. Where $ L_j = \lambda L_{a}+\varkappa^2
L_{b}+\irm k_{aj}L_{c}+\irm k_{bj}L_{d}+\pi L_{e1}-L_{g} -\sigma
R_{Tc}L_{R1}+\sigma R_{Sc}L_{R2}$.

\subsection{Dispersion relation} Parameters of each from $n$ roll-modes
$\lambda, k_{aj}, k_{bj}, R_{Tc}, R_{Sc}$ are related by the equation:
\begin{equation*}
\begin{split}
&(\lambda+\sigma\varkappa^2)(\lambda+\varkappa^2)(\lambda+\tau\varkappa^2)\\
& \qquad\qquad +\sigma(k^2/\varkappa^2)[R_{Sc}(\lambda+\varkappa^2)
-R_{Tc}(\lambda+\tau\varkappa^2)]=0\,.
\end{split}
\end{equation*}
Here $k^2=k_{aj}^2+k_{bj}^2$, and $\varkappa^2=k^2+\pi^2\,.$ This
equation has three roots, two of which can be complex conjugates. In
the case of Hopf bifurcation these two roots acquire positive real
part at some $R_{Tc}$  ($\omega$ is a frequency of convective
waves):
\begin{equation}\label{ra}
\begin{split}
& R_{Tc} = \frac{\sigma+\tau}{1+\sigma}R_{Sc} +
\frac{\varkappa^6}{\sigma k^2}(1+\tau)(\tau+\sigma)\,,\\
& \omega^2 = \frac{1-\tau}{1+\sigma}\sigma
R_{Sc}\frac{k^2}{\varkappa^2}-\tau^2\varkappa^4 > 0\,.
\end{split}
\end{equation}
Here $\omega$ is a frequency of convective waves, and it is assumed
to be real. This means that the number $R_{Sc}$ should not be too small.
In this paper we consider double-diffusive convection at Hopf bifurcation points,
i.e. in all cases $\lambda = \irm\omega$.

\subsection{Critical wavenumber} From the expressions (\ref{ra}) one can see that
the minimal Rayleigh number $R_{Tc}$ is obtained at
$k_c=\pi/\sqrt{2}$, which defines the characteristic size of
convective cells, arising with an increase of $R_{T}$ above the
critical value. Along with the mode having the wavenumber $k_c$ the
adjacent modes different from the central mode by an amount
$\varepsilon$ also are exited, which leads to the result that the
total wavepackage looks like one mode with the wavenumber $k_c$ and
variable amplitude described by the amplitude equations.

For the sufficiently large Rayleigh numbers the situation is
changing so that the characteristic critical wavenumber is of the
order $0.23\sqrt{\omega}$ and may reach values of $k_c\approx
100$~\cite{KABC}. As in the case of small $R_{S}$ the first losing
stability mode is the mode with $k_c=\pi/\sqrt{2}$. However, with
the growth of $\varepsilon$ the wavenumber of the fastest growing
mode increases proportionally $\sqrt{\varepsilon}$. For some
$\varepsilon$ this growth is stabilized at $k_c\approx 10-100$,
which corresponds to a narrow convective cells. And similarly the
adjacent modes are excited, forming a wave packet, which looks like
one mode with variable amplitude.

So it makes sense to derive the desired amplitude equations for convective
cells of an arbitrary width assuming that the specific value of a small parameter
each time defines the value of $k_c$, which we will further denote as $k$.

\subsection{Shape of the cells}
In the studied system any number of roll modes with different
wavevectors can be excited simultaneously, producing convective
cells of various forms.
Thus, superposition of the two rolls at right angles to each other
gives the square-type cells, three rolls at angles of 120 degrees
form hexagonal cells.
In this paper we do not limit ourselves to any one cell shape, but
consider the general situation, when $n$ roll modes at arbitrary
angles to each other are excited.
The desired amplitude equations will give the opportunity to find
out which of modes given initially in some region of space become
dominant and determine the final shape of the cells.

\section{Resolution conditions}

\subsection{General structure of equations}

The obtained systems have the following general form:
$$
\widehat{L}\varphi_i = Q_i\,.
$$
Functions $Q_i$ include terms, resonating with the left parts of
equations, i.e. $Q_i=Q_i^{(1)}+Q_i^{(2)}+Q_i^{(3)}$. Here
$Q_i^{(1)}$ and $Q_i^{(2)}$ generate the secular terms of two types
in the solutions, but $Q_i^{(3)}$ does'nt generate any secular terms
and contains only unimportant terms for the explored case. The
conditions of the first type secular terms absence reduce to demand
of orthogonality functions $Q_i^{(1)}$ and solutions $F_j$ of the
adjoint homogeneous equation $\widehat{L}^{\star}F_j = 0$ and
usually take form of amplitude equations.
Terms $Q_i^{(2)}$ are the constants with respect to quick variables.
Not to brake the regularity of the asymptotic expansions
(\ref{nmode1a}) they should be equal to zero $Q_i^{(2)}=0$
\cite{BB}. These conditions also take form of amplitude equations.

\subsection{Scalar products} Introduce scalar product of the vectors, composed
of the dependent variables:
\begin{equation*}
\begin{split}
&\langle\varphi_i,\varphi_j\rangle_0 =
\lim_{l\to\infty}\frac{2}{l^3}\int_0^1\left[\iiint_{-l/2}^{l/2}
(u_iu_j+v_iv_j+w_iw_j \right.\\
& \qquad\qquad\qquad\qquad\left.+\theta_i\theta_j+\xi_i\xi_j+p_ip_j) dx dy dt\right] dz\,.
\end{split}
\end{equation*}
The actual forms of the functions $\varphi_i$ and $\varphi_j$,
arising in the explored cases can be the following:
\begin{equation*}
\begin{split}
&\varphi_i=\overline{\varphi}_i\ex^{\irm n_i\omega t}\exi{(\vec{k}_i,\vec{x})}\sico{\pi m_i z}+\cj\,,\\
&\varphi_j=\overline{\varphi}_j\ex^{\irm n_j\omega t}\exi{(\vec{k}_j,\vec{x})}\sico{\pi m_j z}+\cj\,.
\end{split}
\end{equation*}
Then we get:
\begin{equation*}
\begin{split}
&\langle\varphi_i,\varphi_j\rangle_0 = \Ds(\vec{k}_i-\vec{k}_j)\delta_{n_in_j}\delta_{m_im_j}
\langle\overline{\varphi}_i,\overline{\varphi}_j\rangle+\\
&\Ds(\vec{k}_i+\vec{k}_j)\delta_{-n_in_j}\delta_{m_im_j}
\langle\overline{\varphi}_i^*,\overline{\varphi}_j\rangle
+\cj = \langle\varphi_i,\varphi_j\rangle_c +\cj\,.
\end{split}
\end{equation*}
Here as $\delta_{ij}$ we denoted the Kronecker delta, and function
$\Ds(x)$ is defined as $\Ds(0)=1$ and $\Ds(x)=0$ at $x\ne 0$. Also
we have introduced scalar product for the amplitudes of vectors of
the dependent variables:
$$
\langle\overline{\varphi}_i,\overline{\varphi}_j\rangle = \overline{u}_i\overline{u}_j^*+\overline{v}_i\overline{v}_j^*
+\overline{w}_i\overline{w}_j^*+\overline{\theta}_i\overline{\theta}_j^*+\overline{\xi}_i\overline{\xi}_j^*
+\overline{p}_i\overline{p}_j^*\,.
$$
Thus for the sake of amplitude equations derivation from the
resolution conditions we have introduced the three cases of scalar
products: $\langle\varphi_i,\varphi_j\rangle_0$,
$\langle\varphi_i,\varphi_j\rangle_c$ and
$\langle\overline{\varphi}_i,\overline{\varphi}_j\rangle$. The first
one is the initial scalar product in the integral form, the second
and third are introduced for getting the nonlinear and linear terms
of the amplitude equations respectively.

\subsection{The resolution condition} The functions in the right parts of
equations $Q_i^{(1)}$ have the following general form:
$$
Q_i^{(1)}=\sum_{q=1}^{p}\overline{Q}^{(1)}_{iq}\ex^{\irm n_q\omega t}
\exi{(\vec{k}_q,\vec{x})}\sico{\pi m_q z}+\cj\,.
$$
Here $p$ is the number of terms of the considered type in the
functions $Q_i^{(1)}$. The resolution conditions in this case have
the form:
\begin{equation}\label{cmp1}
\begin{split}
&\langle Q_i^{(1)},F_j\rangle_c =\sum_{q=1}^{p}\left[\Ds(\vec{k}_q-\vec{k}_j)\delta_{n_q 1}\delta_{m_q 1}
\langle\overline{Q}_{iq}^{(1)},\overline{F}_j\rangle\right. \\
&\qquad\qquad\left.+\Ds(\vec{k}_q+\vec{k}_j)\delta_{-n_q 1}\delta_{m_q 1}
\langle\overline{Q}_{iq}^{(1)*},\overline{F}_j\rangle\right]=0\,.
\end{split}
\end{equation}
Here we have used the explicit expressions for the vectors $F_j$ of
homogeneous adjoint equation solutions:
$$
F_j=\overline{F}_j\ex^{\irm\omega t}\exi{(\vec{k}_j,\vec{x})}\sico{\pi z}+\cj\,,\quad (L_j^*)^T\overline{F}_j=0\,.
$$
In many cases we can explicitly resolve Kronecker deltas in
equations (\ref{cmp1}), when remain only nonzero terms with
$\vec{k}_q=\vec{k}_j$ and $n_q=m_q=1$. Then the resolution
conditions for the considered systems of equations will
be~\cite{Nayfeh}: $\langle\overline{Q}_{ij},\overline{F}_j\rangle =
0$. I.e. for the compatibility of the obtained algebraic systems of
equations its right parts should be orthogonal to the solutions of
the adjoint homogeneous system. The actual form of the vectors
$\overline{F}_j$ in our case is:
\begin{equation*}
\begin{split}
& \overline{F}_j=\\
&\left(\irm k_{aj}\pi,\irm k_{bj}\pi,k^2,\frac{k^2\sigma R_{Tc}}{\lambda^*+\varkappa^2},
-\frac{k^2\sigma R_{Sc}}{\lambda^*+\tau\varkappa^2},\pi(\lambda^*+\sigma\varkappa^2)\right)^T\,.
\end{split}
\end{equation*}

\section{Equations at $\varepsilon^2$}

\subsection{General frame of derivation}
Find the amplitude equations derived from the system at
$\varepsilon^2$. Write vector of the right parts $Q_2$ as a sum of
three components mentioned earlier:
$$
\Ls\varphi_2 = - \Ls_1\varphi_1-\Ns_1(\varphi_1,\varphi_1)= Q_2^{(1)}+Q_2^{(2)}+Q_2^{(3)}\,.
$$
Note that $\varphi_1=\varphi_1^{(0)}+\widehat{\varphi}_1$, where
$\varphi_1^{(0)}$ is the solution of the homogeneous equation
$\widehat{L}\varphi_1^{(0)}=0$, $\widehat{\varphi}_1$ is the
averaged fields depending only on slow variables. Then write the
expressions for components $Q_2$, omitting zero terms:
\begin{equation*}
\begin{split}
&{Q}_2^{(1)} = - \Ls_1\varphi_1^{(0)}-\Ns_1(\widehat{\varphi}_1,\varphi_1^{(0)})\,,\\
&{Q}_2^{(2)} = - \Ls_1\widehat{\varphi}_1\,,\\
&{Q}_2^{(3)} = -\Ns_1(\varphi_1^{(0)},\varphi_1^{(0)})\,.
\end{split}
\end{equation*}

\subsection{Stream function} To exclude secular terms of the second
type one should require fulfillment of equality ${Q}_2^{(2)} = 0$.
Written in components it gives the following system:
\begin{equation*}
\begin{split}
&{Q}_2^{(2)}(1) = -\widehat{u}_{1T_1}-\widehat{p}_{1X}=0\,,\\
&{Q}_2^{(2)}(2) = -\widehat{v}_{1T_1}-\widehat{p}_{1Y}=0\,,\\
&{Q}_2^{(2)}(6) = -\widehat{u}_{1X}-\widehat{v}_{1Y} = 0\,.
\end{split}
\end{equation*}
To satisfy these equalities introduce horizontal stream function
$\Psi$ by formulas:
\begin{equation}\label{aeq2b}
\widehat{u}_1 = \Psi_{Y},\qquad \widehat{v}_1 = -\Psi_{X}\,, \qquad
\Psi_{T_1}=0\,.
\end{equation}
Also it is true $\widehat{p}_{1}=0$ with the accuracy to constants
on horizontal variables.

\subsection{Amplitude equations} Calculations show that for
$Q_2^{(1)}$ is true the following expression:
\begin{equation*}
\begin{split}
&Q_2^{(1)} = \sum_{j=1}^n\{\check{Q}_{2ja}[A_{jT_1}+(\irm k_{aj}\Psi_Y-\irm k_{bj}\Psi_X)A_j]\\
&\qquad\qquad+\check{Q}_{2jb}A_{jX}+\check{Q}_{2jc}A_{jY}\}\exi{\phi_j}\sico{\pi z}+\cj\,.
\end{split}
\end{equation*}
Here we have introduced phases $\phi_j=\omega t+k_{aj}x+k_{bj}y$ of each mode,
and components of the vectors in the expression are:
\begin{equation*}
\begin{split}
&\check{Q}_{2ja} = - L_a\check{\varphi}_{1j}\,,\qquad
\check{Q}_{2jb} = (2\irm k_{aj}L_b-L_c)\check{\varphi}_{1j}\,,\\
&\check{Q}_{2jc} = (2\irm k_{bj}L_b-L_d)\check{\varphi}_{1j}\,.
\end{split}
\end{equation*}
The condition of there be no secular terms of the first type (\ref{cmp1}) in
the solutions of the equations at $\varepsilon^2$ is written as $\langle
Q_2^{(1)},F_j\rangle_c = 0$ and, after some calculations, it reduces to
requirement
$\langle\overline{Q}_{2j}^{(1)},\overline{F}_j\rangle = 0$ for
each $j=1\ldots n$. Or more explicitly:
\begin{equation*}
\begin{split}
&\langle\overline{Q}_{2j}^{(1)},\overline{F}_j\rangle =
\langle\check{Q}_{2ja},\overline{F}_j\rangle [A_{jT_1}+(\irm
k_{aj}\Psi_Y-\irm k_{bj}\Psi_X)A_j]\\
& \qquad\qquad\qquad+\langle\check{Q}_{2jb},\overline{F}_j\rangle
A_{jX}+ \langle\check{Q}_{2jc},\overline{F}_j\rangle A_{jY}=0\,.
\end{split}
\end{equation*}
Finally the amplitude equations take the following form:
\begin{equation}\label{aeq2a}
\begin{split}
&A_{jT_1}+2\alpha_0(\irm k_{aj}A_{jX}+\irm k_{bj}A_{jY})\\
&\qquad+(\irm k_{aj}\Psi_Y-\irm k_{bj}\Psi_X)A_j=0\,,\quad j=1\ldots n\,.
\end{split}
\end{equation}
Where $\alpha_0=\langle\check{Q}_{2jb},\overline{F}_j\rangle/(2\irm
k_{aj}\langle\check{Q}_{2ja},\overline{F}_j\rangle)\,,$ or finally:
\begin{equation}\label{alf0}
\begin{split}
&\alpha_0 =
\frac{\irm\omega}{\varkappa^2}\left[1+\left(\frac{\pi^2}{2k^2}-1\right)\times\right.\\
&\qquad\times\left.\left(1-\frac{\varkappa^4}{\omega^2} \cdot
\frac{(\tau+\sigma+\tau\sigma)\irm\omega+\tau\sigma\varkappa^2}%
{\irm\omega+(1+\tau+\sigma)\varkappa^2}\right)\right]\\
&\qquad\qquad=\frac{\irm\omega}{\varkappa^2}+\beta\,.
\end{split}
\end{equation}
Here we have introduced coefficient $\beta$, which is evidently
defined by the above expression. Equations (\ref{aeq2b}) in many
important cases can be resolved explicitly and usually imply some
kind of transport, so further we don't discuss their solutions.

Equations (\ref{aeq2b}) and (\ref{aeq2a}) together consist the
desired system of amplitude equations obtained as a result of
consideration of the members at $\varepsilon^2$ in the
multiple-scaled method. If the first one is satisfied by introducing
a horizontal stream function $\Psi(X,Y,T_2)$ independent on the slow
time $T_1$, then the second one will be used to exclude members
alike $A_{jT_1}$ from the final amplitude equations.
Obtained for $A(X,Y,T_2)$ and $\Psi(X,Y,T_2)$ solutions of equations
(\ref{eq_fin}) one should substitute into the equations
(\ref{aeq2a}) to find the dependence of the amplitudes from $T_1$.

\section{Equations at $\varepsilon^3$}

\subsection{General frame of derivation}
At last we write the resulting family of amplitude equations for the system at
$\varepsilon^3$. For this purpose we need the solutions for
$\varphi_1$ and $\varphi_2$, which can be expressed in a general form:
$$
\varphi_1=\varphi_1^{(0)}+\widehat{\varphi}_1\,,\qquad
\varphi_2=\varphi_2^{(0)}+\varphi_2^{(1)}+\widehat{\varphi}_2+\widetilde{\varphi}_2\,.
$$
Here $\varphi_i^{(0)}$ are the general solutions of homogeneous equations
$\widehat{L}\varphi_i^{(0)}=0$, $\varphi_2^{(1)}$ and
$\widetilde{\varphi}_2$ are linear and nonlinear on amplitude terms of the
particular solution of the inhomogeneous equation
$\widehat{L}(\varphi_2^{(1)}+\widetilde{\varphi}_2)=Q_2$,
$\widehat{\varphi}_i$ are the averaged fields on the slow horizontal equations, arising as an
integrating constants. For $\varphi_2^{(1)}$ we have the following expression:
\begin{equation*}
\begin{split}
& \varphi_2^{(1)} =
\sum_{j=1}^n(\check{\varphi}_{2jg}A_{jX}+\check{\varphi}_{2jh}A_{jY})\exi{\phi_j}\sico{\pi z} + \cj\,,\\
%\qquad
&\check{\varphi}_{2jg} = \check{\varphi}_{2jb}-2\alpha_0\irm k_{aj}\check{\varphi}_{2ja}\,,
 \check{\varphi}_{2jh}=(\check{\varphi}_{2jc}-2\alpha_0\irm k_{bj}\check{\varphi}_{2ja})\,,\\
%\quad
&\text{where}\quad
L_j\check{\varphi}_{2ja}=\check{Q}_{2ja}\,,\quad
L_j\check{\varphi}_{2jb}=\check{Q}_{2jb}\,,\quad
L_j\check{\varphi}_{2jc}=\check{Q}_{2jc}\,.
\end{split}
\end{equation*}

Write the system at $\varepsilon^3$ in a general form:
\begin{equation*}
\begin{split}
&\Ls\varphi_3 = - \Ls_1\varphi_2-\Ls_2\varphi_1-\Ns_1(\varphi_1,\varphi_2)-\Ns_1(\varphi_2,\varphi_1)\\
&\qquad\qquad-\Ns_2(\varphi_1,\varphi_1)= Q_3^{(1)}+Q_3^{(2)}+Q_3^{(3)}\,.
\end{split}
\end{equation*}
Then the expressions for $Q_3^{(1)}$ and $Q_3^{(2)}$, the only needed for the derivation of amplitude equations
take form:
\begin{equation*}
\begin{split}
& Q_3^{(1)} =-[\Ls_1\varphi_2^{(1)}+\Ls_2\varphi_1^{(0)}+
\Ns_1(\widehat{\varphi}_1,\varphi_2^{(1)})
+\Ns_2(\widehat{\varphi}_1,\varphi_1^{(0)})\\
& \qquad\qquad +\Ns_2(\varphi_1^{(0)},\widehat{\varphi}_1)+\Ns_1(\varphi_1^{(0)},\widetilde{\varphi}_2)
+\Ns_1(\widetilde{\varphi}_2,\varphi_1^{(0)})]\\
&\qquad\qquad\qquad\qquad\qquad\qquad=Q_{3}^{(1l)}+Q_{3}^{(1p)}+Q_{3}^{(1n)}\,,\\
& Q_3^{(2)} =
-[\Ls_1\widehat{\varphi}_2+\Ls_2\widehat{\varphi}_1+\Ns_1(\varphi_2^{(1)},\varphi_1^{(0)})\\
&\qquad\qquad\qquad\qquad+\Ns_2(\varphi_1^{(0)},\varphi_1^{(0)})+\Ns_2(\widehat{\varphi}_1,\widehat{\varphi}_1)]\,.
\end{split}
\end{equation*}
Here we have separately identified linear $Q_{3}^{(1l)}$ and nonlinear $Q_{3}^{(1n)}$ on $A_j$ terms,
and also terms $Q_{3}^{(1p)}$, containing $\widehat{\varphi}_1$.
Denote $\beta_0=-\langle\check{Q}_{2ja},\overline{F}_j\rangle$, then the desired amplitude equations
in a general form are:
\begin{equation*}
\begin{split}
& \frac{1}{\beta_0}\langle{Q}_{3}^{(1)},{F}_j\rangle_c =
\frac{1}{\beta_0}\langle\overline{Q}_{3}^{(1l)},\overline{F}_j\rangle\\
&\qquad\qquad\qquad\qquad+\frac{1}{\beta_0}\langle\overline{Q}_{3}^{(1p)},\overline{F}_j\rangle+
\frac{1}{\beta_0}\langle{Q}_{3}^{(1n)},{F}_j\rangle_c=0\,.
\end{split}
\end{equation*}

\subsection{Linear terms}
The detailed calculations give the following formula for the linear terms of equations:
\begin{equation*}
\begin{split}
& \frac{1}{\beta_0}\langle\overline{Q}_{3}^{(1l)},\overline{F}_j\rangle=-A_{jT_2}
+\frac{1}{\beta_0}\langle\check{Q}_{3jR},\overline{F}_j\rangle A_j\\
&\qquad+\frac{1}{\beta_0}\langle\check{Q}_{3ja},\overline{F}_j\rangle A_{jXX}
+\frac{1}{\beta_0}\langle\check{Q}_{3jb},\overline{F}_j\rangle A_{jYY}+\\
&\qquad +\frac{1}{\beta_0}\langle\check{Q}_{3jc},\overline{F}_j\rangle A_{jXY}=
-A_{jT_2}+rA_j-\alpha_0\Delta_{\perp}A_j\\
&\qquad\qquad+\frac{\alpha_1}{k^2}(k_{aj}^2 A_{jXX}+2k_{aj}k_{bj}A_{jXY}+k_{bj}^2A_{jYY})\,.
\end{split}
\end{equation*}

\subsection{Terms with stream function}
Similarly one can get a formula for the terms with $\Psi$:
\begin{equation*}
\begin{split}
& \frac{1}{\beta_0}\langle\overline{Q}_{3}^{(1p)},\overline{F}_j\rangle=\\
&\qquad-\frac{1}{\beta_0}\left(\frac{1}{\irm k_{aj}}\langle\check{Q}_{3jd},\overline{F}_j\rangle
+\frac{\pi^2}{k^2}\right)A_j[(k_{aj}^2-k_{bj}^2)\Psi_{XY}\\
&\qquad+k_{aj}k_{bj}(\Psi_{YY}-\Psi_{XX})]-J(A_j,\Psi)=J(\Psi, A_j)+\\
&\qquad+\frac{\irm k\alpha_3}{k^2}A_j[(k_{aj}^2-k_{bj}^2)\Psi_{XY}+k_{aj}k_{bj}(\Psi_{YY}-\Psi_{XX})]\,.
\end{split}
\end{equation*}
Here the Jacobian $J(f,g)=f_Xg_Y-f_Yg_X$ is introduced.
In the formulas the following vectors of the right parts of equations are used:
\begin{equation*}
\begin{split}
& \check{Q}_{2ja}=-L_{a}\check{\varphi}_{1j}\,,\qquad\check{Q}_{3jd}=L_{a}\check{\varphi}_{2jg}\,,\\
& \check{Q}_{3jR} = -\sigma(r_T R_{Tc}L_{R1}-r_S R_{Sc}L_{R2})\check{\varphi}_{1j}\,,\\
%&\check{Q}_{3jd}=L_{a}\check{\varphi}_{2jg}\,,\\
& \check{Q}_{3ja}=(2\alpha_0\irm k_{aj}L_{a}
+2\irm k_{aj}L_{b}-L_{c})\check{\varphi}_{2jg}+L_{b}\check{\varphi}_{1j}\,,\\
& \check{Q}_{3jb}=(2\alpha_0\irm k_{bj}L_{a}
+2\irm k_{bj}L_{b}-L_{d})\check{\varphi}_{2jh}+L_{b}\check{\varphi}_{1j}\,,\\
& \check{Q}_{3jc}=(2\alpha_0\irm k_{bj}L_{a}+2\irm k_{bj}L_{b}-L_{d})\check{\varphi}_{2jg}\\
&\qquad +(2\alpha_0\irm k_{aj}L_{a}+2\irm k_{aj}L_{b}-L_{c})\check{\varphi}_{2jh}\,.
\end{split}
\end{equation*}

\subsection{Nonlinear terms} Now calculate nonlinear terms
$\langle{Q}_{3}^{(1n)},{F}_j\rangle_c$, in the amplitude equations.
Vector of nonlinear on amplitude members ${Q}_{3}^{(1n)}$ in the right part of the equations
at $\varepsilon^3$ one can represent as the sum:
$$
Q_{3}^{(1n)}=-\left[\Ns_2(\varphi_1^{(0)},\widetilde{\varphi}_2)
+\Ns_2(\widetilde{\varphi}_2,\varphi_1^{(0)})\right]\,.
$$
Scrupulous and quite cumbersome calculation of this sum consist in
that the term $\widetilde{\varphi}_2$ is splited into six terms,
representing the nonlinear terms of the different types. Then it is
substituted into the expression for $Q_{3}^{(1n)}$. After that we
exclude obviously zero terms and remaining five terms of the desired
sum are calculated successively. Their substitution into the
expression for the resolution condition gives for the nonlinear
terms of the amplitude equations the following final formulas:
\begin{equation*}
\begin{split}
& \frac{1}{\beta_0}\langle\overline{Q}_{3}^{(1n)},\overline{F}_j\rangle_c =
\alpha_2A_j\sum_{q=1}^{n}|A_q|^2\\
&+\sum_{m=1}^n\sum_{q=1}^{n}\sum_{p=q+1}^n\left[
\Ds(\vec{k}_q+\vec{k}_p-\vec{k}_m-\vec{k}_j)\alpha_{jmqp}^{(1)}A_m^*A_qA_p\right. \\
&\qquad\qquad+\Ds(\vec{k}_q-\vec{k}_p-\vec{k}_m+\vec{k}_j)\alpha_{jmqp}^{(2)}A_mA_q^*A_p\\
&\qquad\qquad\left.+\Ds(\vec{k}_q-\vec{k}_p+\vec{k}_m-\vec{k}_j)\alpha_{jmqp}^{(3)}A_mA_qA_p^*\right]\,.
\end{split}
\end{equation*}
For the coefficients $\alpha_{jmqp}^{(s)}, (s=1,2,3)$, the following
expressions are true:
\begin{equation}\label{alf_n}
\begin{split}
& \alpha_{jmqp}^{(1)} =
-\frac{c_{qp1}\pi^2}{2k^4}\beta_2\left\{\left[\left(1-\frac{c_{mqp1}}{c_{qp2}}\right)
\frac{\pi^2}{\varkappa^2}c_{jm1}\right.\right.\\
&\left.\left.+(2k^2-c_{mqp1})\beta_{15}\right]\beta_{4}+(2k^2-c_{mqp1})\beta_7\right\},\\
& \alpha_{jmqp}^{(2)} =
-\frac{c_{qp2}\pi^2}{2k^4}\left\{\left[\left(1-\frac{c_{mqp2}}{c_{qp1}}\right)\beta_{11}\right.\right.\\
&\left.\left.+(2k^2-c_{mqp2})\left(\beta_{12}-\frac{\pi^2 c_{jqp2}}{k^2c_{qp1}}\right)
\frac{k^2}{\varkappa^2}\beta_2\right]\beta_{5}-\beta_{14}\right\},\\
& \alpha_{jmqp}^{(3)} =
-\frac{c_{qp2}\pi^2}{2k^4}\left\{\left[\left(1+\frac{c_{mqp2}}{c_{qp1}}\right)\beta_{11}\right.\right.\\
&\left.\left.+(2k^2+c_{mqp2})\left(\beta_{12}+\frac{\pi^2 c_{jqp2}}{k^2
c_{qp1}}\right)\frac{k^2}{\varkappa^2}\beta_2\right]\beta_{5}
-\beta_{13}\right\}.
\end{split}
\end{equation}
Here coefficients $\beta_4, \beta_5, \beta_6, \beta_7$, depending from
$p$ and $q$ are denoted:
\begin{equation*}
\begin{split}
& \beta_{4} = \left[1-\Ds(\vec{k}_q+\vec{k}_p)\right]\times\\
&\times\left.\left\{2+\frac{(c_{qp2}+2\pi^2)\beta_8+\irm\omega(\irm\omega+\sigma\varkappa^2)}%
{(\irm\omega+c_{qp2}+2\pi^2)(\irm\omega+\tau(c_{qp2}+2\pi^2))}\right\}\right/ \\
& \qquad \left/\left\{\frac{4k^2(c_{qp2}+2\pi^2)(\irm\omega+\sigma(c_{qp2}+2\pi^2))}{c_{qp2}\varkappa^2}+\right.\right.\\
&\qquad\qquad\left.\left.+\frac{(c_{qp2}+2\pi^2)\beta_9+\irm\omega\beta_{10}}%
{(\irm\omega+c_{qp2}+2\pi^2)(\irm\omega+\tau(c_{qp2}+2\pi^2))}\right\}\right.\,,
\end{split}
\end{equation*}
\begin{equation*}
\begin{split}
& \beta_{5} =\frac{c_{qp1}\tau\varkappa^2(2c_{qp1}+4\pi^2-\varkappa^2)}
{4k^2\tau\sigma(c_{qp1}+2\pi^2)^3+c_{qp1}\varkappa^2\beta_9}\,,\\
& \beta_6 = \frac{(c_{qp2}+2\pi^2)\beta_8+\irm\omega(\irm\omega+\sigma\varkappa^2)}%
{2(\irm\omega+c_{qp2}+2\pi^2)(\irm\omega+\tau(c_{qp2}+2\pi^2))} \,,\\
\end{split}
\end{equation*}
\begin{equation*}
\begin{split}
& \beta_7 = \left\{(c_{qp2}+2\pi^2)[(\irm\omega+\varkappa^2)\beta_8
+\tau\varkappa^2(\tau+\sigma)(\irm\omega-\varkappa^2)]\right.\\
&\qquad\qquad\qquad\qquad\left.\left.+\irm\omega\varkappa^2\beta_8-\omega^2(\irm\omega+\sigma\varkappa^2)\right\}\right/\\
&\qquad\qquad\left/\left\{2(\irm\omega+c_{qp2}+2\pi^2)(\irm\omega+\tau(c_{qp2}+2\pi^2))\right.\right.\\
&\qquad\qquad\qquad\qquad\left.\times(\irm\omega+\varkappa^2)(\irm\omega+\tau\varkappa^2)\right\}\,.
\end{split}
\end{equation*}
For a more compact form of the formulas the coefficients
$\beta_8 =
(1+\tau+\sigma)\irm\omega-\tau\varkappa^2$ and also $\beta_9 =
(1+\tau+\sigma)\omega^2-\tau\sigma\varkappa^4$ and $\beta_{10} =
\omega^2-(\tau+\sigma+\tau\sigma)\varkappa^4$ are introduced.
It is worth to mention that coefficient $\beta_4$ turns to zero
each time when $\vec{k}_q=-\vec{k}_p$ is true.
In addition in formulas (\ref{alf_n}) a few more coefficients are
used:
\begin{equation*}
\begin{split}
& \beta_{11} = \frac{\pi^2}{\varkappa^2}c_{jm2}\beta_2-k^2\,,
\qquad \beta_{12} = 1 +\frac{\varkappa^2\beta_8}{2\tau k^2(c_{qp1}+2\pi^2)}\,,\\
&\beta_{13} = \frac{\varkappa^2(2k^2+c_{mqp2})}{4\irm\omega(c_{qp1}+2\pi^2)}\,,
\qquad \beta_{14} = \frac{\varkappa^2(2k^2-c_{mqp2})}{4\irm\omega(c_{qp1}+2\pi^2)}\,,\\
&\beta_{15} = \left(1+\frac{\pi^2 c_{jqp1}}{k^2 c_{qp2}}\right)\frac{k^2}{\varkappa^2}+\beta_6\,.
\end{split}
\end{equation*}
Here $\beta_2$ is defined by the formula (\ref{bet2}). In all
represented above formulas the values, composed from the scalar
products of the mode wavenumbers are used:
$c_{mqp1}=(\vec{k}_m,\vec{k}_q)+(\vec{k}_m,\vec{k}_p)$,
$c_{mqp2}=(\vec{k}_m,\vec{k}_q)-(\vec{k}_m,\vec{k}_p)$,
$c_{qp1}=k^2-(\vec{k}_q,\vec{k}_p)$,
$c_{qp2}=k^2+(\vec{k}_q,\vec{k}_p)$.

\subsection{Equation for the stream function}
From the condition of there be no secular terms of the second type
in the equations at $\varepsilon^3$ one should require
to be true ${Q}_3^{(2)} = 0$. Written in components it
leads to the following system of equations:
\begin{equation*}
\begin{split}
& Q_3^{(2)}(1)=
\sigma\Delta_{\perp}\Psi_{Y}-\Psi_{YT_2}-\Psi_{2YT_1}-\widehat{p}_{2X}
-\Psi_Y\Psi_{XY}\\
&+\Psi_{X}\Psi_{YY}-\frac{\pi^2}{k^4}\sum_{j=1}^n(k_{aj}^2(|A_j|^2)_X+k_{aj}k_{bj}(|A_j|^2)_Y)=0\,,\\
& Q_3^{(2)}(2)=
-\sigma\Delta_{\perp}\Psi_{X}+\Psi_{XT_2}+\Psi_{2XT_1}-\widehat{p}_{2Y}
+\Psi_Y\Psi_{XX}\\
&-\Psi_{X}\Psi_{XY}-\frac{\pi^2}{k^4}\sum_{j=1}^n(k_{aj}k_{bj}(|A_j|^2)_X+k_{bj}^2(|A_j|^2)_Y)=0\,.
\end{split}
\end{equation*}
Differentiate the first of these equations with respect to $Y$
and subtract from it the second equation differentiated with
respect to $X$. Also assume that $\Delta_{\perp}\Psi_{2T_1}=0$,
as it is true in the case of $\Psi$.
As a result we obtain the final equation, relating horizontal
vorticity $\Psi$ with convection  $A_j$.
$$
(\partial_{T_2}-\sigma\Delta_{\perp})\Delta_{\perp}\Psi=J(\Psi,\Delta_{\perp}\Psi)
-\frac{\pi^2}{k^2}\sum_{j=1}^n\widehat{G}_j(|A_j|^2)
$$
Here we have introduced linear operator
$$\widehat{G}_j(f)=\frac{1}{k^2}(k_{aj}\partial_X+k_{bj}\partial_Y)
(k_{aj}\partial_Y-k_{bj}\partial_X)f\,.$$

\section{The $A_j\Psi-$family of amplitude equations}

Finally we write the resulting family of amplitude equations for the
system at $\varepsilon^3$:
\begin{equation}\label{eq_fin}
\begin{split}
%\begin{cases}
&
\partial_{T_2}A_j=rA_j+\displaystyle{\frac{\alpha_1}{k^2}}(k_{aj}\partial_X+k_{bj}\partial_Y)^2A_j
-\alpha_0\Delta_{\perp}A_j\\
&+\irm k\alpha_3\widehat{G}_j(\Psi)A_j+J(\Psi,A_j)+N_j(A)\,,\\
&(\partial_{T_2}-\sigma\Delta_{\perp})\Omega=J(\Psi,\Omega)
-\displaystyle{\frac{\pi^2}{k^2}\sum_{j=1}^n}\widehat{G}_j(|A_j|^2)\,,\\
& \Omega=\Delta_{\perp}\Psi\,.
%\quad j=1\ldots n\,.
%\end{cases}
\end{split}
\end{equation}
Where $\Delta_{\perp}$ is Laplacian with respect to the slow variables, $\alpha_i$ are complex
coefficients. Index $j=1\ldots n$  denotes the mode number.
This family of the systems of amplitude equations depends on the set of n
wavevectors which define the shape of convective cells. Operator $\widehat{G}$
in the equations describes an interaction between convection and
field of horizontal vorticity, generation of vortex due to convection.

The functions $N_j(A)$ are the following combination of cubic nonlinear terms:
\begin{equation}
\begin{split}
& N_j(A)=
\alpha_2A_j\sum_{q=1}^{n}|A_q|^2\\
& +\sum_{m=1}^n\sum_{q=1}^{n}\sum_{p=q+1}^n\left[
\Ds(\vec{k}_q+\vec{k}_p-\vec{k}_m-\vec{k}_j)\alpha_{jmqp}^{(1)}A_m^*A_qA_p+\right. \\
&+\Ds(\vec{k}_q-\vec{k}_p-\vec{k}_m+\vec{k}_j)\alpha_{jmqp}^{(2)}A_mA_q^*A_p\\
&+\left.\Ds(\vec{k}_q-\vec{k}_p+\vec{k}_m-\vec{k}_j)\alpha_{jmqp}^{(3)}A_mA_qA_p^*
\right]\,.
\end{split}
\end{equation}
Coefficients in these equations are defined by expressions:
\begin{equation}\label{bet}
\begin{split}
& r =
\beta_2\frac{(\sigma+\tau)(\varkappa^2-\irm\omega)r_T-(\sigma+1)(\tau\varkappa^2-\irm\omega)r_S}
{(1-\tau)}\,,\\
& \alpha_1 =
\left(\frac{\pi^2}{k^2}-1\right)\left(\frac{2\irm\omega}{\varkappa^2}+\frac{2\varkappa^2}{\irm\omega}\beta_1\right)-\\
&\qquad-\frac{8k^2}{\varkappa^2}\beta\left[1+\left(\frac{\pi^2}{2k^2}-1\right)\beta_2\right]+\frac{4k^2}{\varkappa^2}\beta^2\beta_3\,,\\
& \alpha_2 = \frac{\varkappa^2}{4\irm\omega}\,,\quad
 \alpha_3 = \frac{2\irm k}{\varkappa^2}\left[1+\left(\frac{\pi^2}{k^2}-1\right)\beta_2-\beta\beta_3\right]\,.
\end{split}
\end{equation}
Here for the convenience and compactness of the expressions we introduce functions:
\begin{equation}\label{bet2}
\begin{split}
& \beta_1 =
\frac{(\tau+\sigma+\tau\sigma)\irm\omega+\tau\sigma\varkappa^2}
{\irm\omega+(1+\tau+\sigma)\varkappa^2}\,, \\
& \beta_2 = \frac{(\irm\omega+\varkappa^2)(\irm\omega+\tau\varkappa^2)}
{2\irm\omega(\irm\omega+(1+\tau+\sigma)\varkappa^2)}=\frac{\varkappa^2}{\beta_0} \,,\\
& \beta_3 =
\frac{\varkappa^2(\omega^2+\tau\varkappa^4)}{2\irm\omega(\irm\omega+\varkappa^2)(\irm\omega+\tau\varkappa^2)}\\
&\qquad-\frac{\varkappa^4((1+\tau+2\sigma)\irm\omega+(1+\tau^2+\tau\sigma+\sigma)\varkappa^2)}{2(\irm\omega+\varkappa^2)
(\irm\omega+\tau\varkappa^2)(\irm\omega+(1+\tau+\sigma)\varkappa^2)}\,.\\
\end{split}
\end{equation}
Coefficients $\alpha_0$ and $\beta$ are given by the formula (\ref{alf0}).
Coefficients $r\,,\alpha_0\,,\alpha_1\,,\alpha_2\,,\alpha_3$ in the equations (\ref{eq_fin}) coincide
with the same-named coefficients in the article \cite{K15}. Therein one can find the expressions
for these coefficients at $k=\pi/\sqrt{2}$ and graphs of their dependence from frequency $\omega$.
Coefficients at the nonlinear terms $\alpha_{jmqp}^{(s)}, (s=1,2,3)$
are presented by the formulas (\ref{alf_n}).

\section{Special cases of amplitude equations for the cells of different forms}

\subsection{Compatibility with solutions for 2D convection}
If we neglect the interaction with the horizontal stream function
and consider the dynamics on the single spatial variable the
obtained one-mode system reduces to the well known Ginzburg-Landau
equation (CGLE):
$$A_{T_2}= r A+\alpha_5 A_{XX} +\alpha_2 A|A|^2\,.$$
Where $\alpha_5=\alpha_1-\alpha_0$.
In the limit of high Hopf frequencies
the resulting equation reduces to the nonlinear Schr\"odinger
equation (NSE) and has ``dark'' solitons solutions~\cite{K10}.

\subsection{Roll type one-mode convection.}
Consider the one-mode convection with convective rolls placed along
the $x$-axis. The wave vector is: $\vec{k}=(k\,,0)$. In this case
the equations (\ref{eq_fin}) after some transformations of dependent
and independent variables take the following shape:
\begin{equation}\label{eq1ap}
\begin{split}
%\begin{cases}
& A_T=A+\alpha_6 A_{XX}-\alpha_7 A_{YY}+\alpha_9\Psi_{XY}A\\
&\qquad\qquad\qquad\qquad\qquad+J(\Psi,A)-\irm A|A|^2\,,\\
& \Omega_T=\alpha_8\Delta_{\perp}\Omega+J(\Psi,\Omega)-(|A|^2)_{XY}\,,\\
& \Omega=\Delta_{\perp}\Psi\,.\\
%\end{cases}
\end{split}
\end{equation}
Here the new coefficients are:
\begin{equation*}
\begin{split}
& \alpha_6=\alpha_5{k^2\varkappa^2}/{(4\pi^2\omega)}\,, \qquad
\alpha_7=\alpha_0{k^2\varkappa^2}/{(4\pi^2\omega)}\,, \\
& \alpha_8={\sigma k^2\varkappa^2}/{(4\pi^2\omega)}\,, \qquad
\alpha_9=\irm k\alpha_3\irm\,.
\end{split}
\end{equation*}
One should especially note, that in the limit of large $\omega$ the
coefficients $\alpha_6$ and $\alpha_7$ don't vanish and become equal
$\alpha_6=3\irm k^2/(8\varkappa^2)$ and $\alpha_7=\irm/8$. In this
limit at the different values of $k$ it is true
$\alpha_6=0\irm\ldots 3\irm/8$. The coefficient $\alpha_8$,
describing attenuation of the vortex $\Omega$, vanishes.
Nevertheless the cross members describing in the equations
interaction of the vortex and convection don't vanish, as one could
expect. For the coefficient $\alpha_9$ at large $\omega$ it is true
$\alpha_9 \approx -{3\pi^2}/{2\varkappa^2}+{\sigma
k^2}/{\irm\omega}$.
Thus one can assume that for the physical macro systems with
double-diffusive convection, for which the sufficiently large values
of $\omega$ are typical, the effects of interaction of the
convection with the field of horizontal vorticity, excitation of the
vortex due to convection play an essential role.

If we assume $\Psi=0$ in equations (\ref{eq1ap}) and exclude the
forcing term, then the derived system reduces to the one equation,
which is the case of 2D nonlinear Schr\"odinger equation (NSE):
$$\irm A_T=-A_{XX}+A_{YY}+A|A|^2\,.$$
Possibly this equation can play an essential role in modeling of
pattern formation processes in various physical systems and
describes the so called {\it dry turbulence} \cite{Dry}.

\subsection{Hexagonal type three-mode convection}
Consider three-mode convection in the case when the convective rolls
are placed at the angles 120 degrees with respect to each other. The
wave vectors are: $\vec{k_1}=(k\,,0),$
$\vec{k_2}=(-k/2\,,k\sqrt{3}/2),$ $\vec{k_3}=(-k/2\,,-k\sqrt{3}/2).$
The system (\ref{eq_fin}) transforms to the following shape:
\begin{equation*}
\begin{split}
%\begin{cases}
& A_T=A+\alpha_6 A_{XX}-\alpha_7 A_{YY}+J(\Psi,A)+\alpha_9A\Psi_{XY}\\
& \qquad -\irm A|A|^2+\alpha_{11}A(|B|^2+|C|^2)\,,\\
& B_T=B+\alpha_{61}B_{XX}+\alpha_{71}B_{YY}-\alpha_{72}B_{XY}+J(\Psi,B)\\
& \qquad -\frac{1}{2}\alpha_{9}B[\Psi_{XY}+\frac{\sqrt{3}}{2}(\Psi_{YY}-\Psi_{XX})]\\
& \qquad -\irm B|B|^2+\alpha_{11}B(|A|^2+|C|^2)\,,\\
& C_T=B+\alpha_{61}C_{XX}+\alpha_{71}C_{YY}+\alpha_{72}C_{XY}+J(\Psi,C)\\
& \qquad -\frac{1}{2}\alpha_{9}C[\Psi_{XY}-\frac{\sqrt{3}}{2}(\Psi_{YY}-\Psi_{XX})]\\
& \qquad -\irm C|C|^2+\alpha_{11}C(|A|^2+|B|^2)\,,\\
& \Omega_T=\alpha_8\Delta_{\perp}\Omega+J(\Psi,\Omega)-(|A|^2-\frac{1}{2}|B|^2-\frac{1}{2}|C|^2)_{XY}\\
& \qquad +\frac{\sqrt{3}}{4}(|B|^2-|C|^2)_{XX}-\frac{\sqrt{3}}{4}(|B|^2-|C|^2)_{YY}\,,\\
& \Omega=\Delta_{\perp}\Psi\,.\\
%\end{cases}
\end{split}
\end{equation*}
Here we denoted the coefficients: $\alpha_{61}=\frac{1}{4}\alpha_{6}-\frac{3}{4}\alpha_{7}$,
$\alpha_{71}=\frac{3}{4}\alpha_{6}-\frac{1}{4}\alpha_{7}$,
$\alpha_{61}=\frac{\sqrt{3}}{2}(\alpha_{6}+\alpha_{7})$.%,

\section{Numerical experiments}

\subsection{The details of calculation methods}
For numerical simulation of the equations (\ref{eq_fin}) the
software packages based on ETD (exponential time differencing)
pseudo-spectral methods~\cite{Cox} were written to study roll-type
convection and convection with square and hexagonal type cells.

In the calculations we used the numerical schemes developed in the
frames of two-layers method ETD2 and ETD2RK method from~\cite{Cox}.
The number of nodes on both horizontal variables was usually 256.
The size of the area for calculations as a rule was chosen as
$15\times15$, and the calculations were led up to the times about
$T=50$. In some cases a square areas of the sizes $10\times10$ and
$25\times25$ were used. In all cases we used the periodic boundary
conditions natural for the pseudo-spectral methods.

As an initial conditions for simulation we choose either an
arbitrary noise with the amplitude $10^{-4}$, or Gauss bell-like
function $A = 2\exp(-0.5(X^2+Y^2))$.

We have performed numerical simulation for the tree cases of
convection. Parameters and coefficients for these cases are the
following:

Case 1: $\omega = 2000,$ $k_c = 10,$ $\alpha_6 = 0.098+0.263\irm,$
$\alpha_7 = -0.019+0.168\irm,$ $\alpha_8 = 0.974,$ $\alpha_9 =
-0.000197-0.277,$ $\alpha_{11} = 0.018-0.688\irm.$

Case 2: $\omega = 20000,$ $k_c = 32.3,$ $\alpha_6 = 1.043-0.42\irm,$
$\alpha_7 = -0.185+0.564\irm,$ $\alpha_8 = 9.72,$ $\alpha_9 =
0.123-0.293\irm,$ $\alpha_{11} = 0.000658-0.961\irm.$

Case 3: $\omega = 150000,$ $k_c = 62.8,$ $\alpha_6 =
0.656-0.875\irm,$ $\alpha_7 = -0.0993+0.596\irm,$ $\alpha_8 =
18.43,$ $\alpha_9 = 0.0374-0.173\irm,$ $\alpha_{11} =
0.00008-0.989\irm.$

In all cases $\sigma = 7$, $\tau = 1/81.$

\begin{figure}
\onefigure[width=0.49\textwidth]{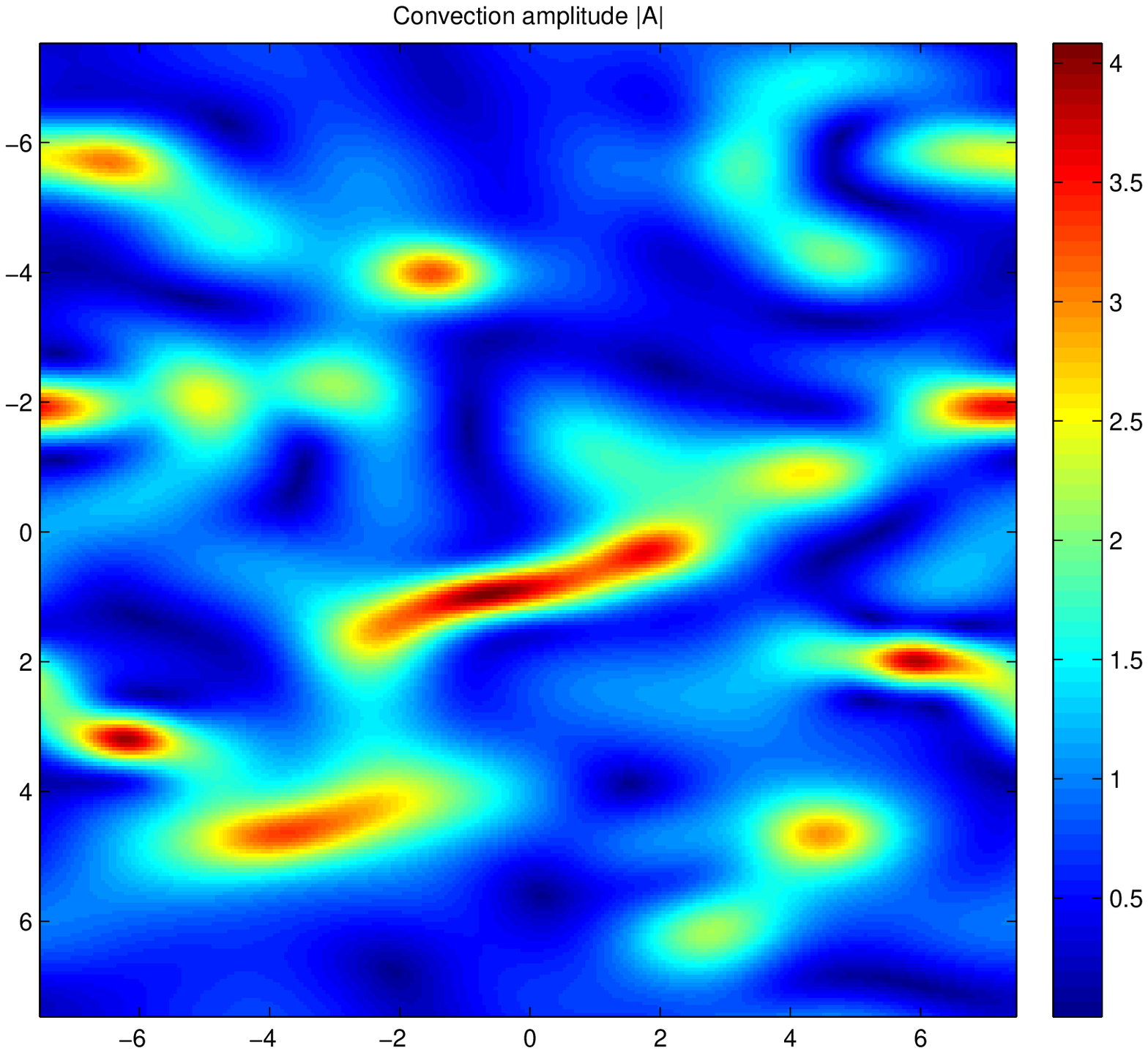}
\caption{Numerical solution of the one-mode equations in the
area $15\times 15$ for $T=20$ at $\omega=20000$.
On the figure the modulus of the amplitude $|A(T,X,Y)|$ is
represented. Initial conditions are an arbitrary noise of
the amplitude $10^{-4}$.}
\label{fig1}
\end{figure}

\begin{figure}
\onefigure
[width=0.49\textwidth]{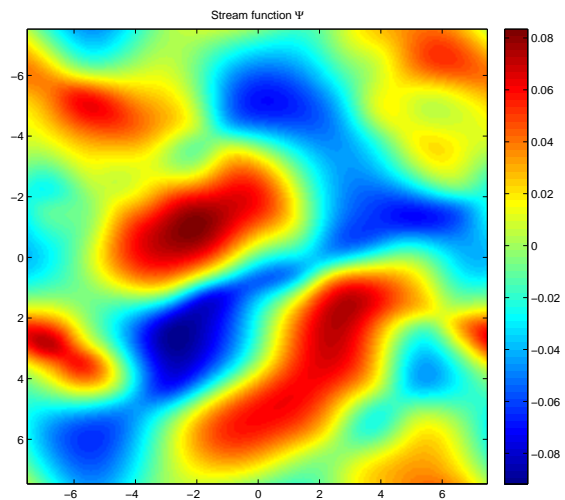}
\caption{Numerical solution of the one-mode equations in the
area $15\times 15$ for $T=20$ at $\omega=20000$.
On the figure the stream function $\Psi(T,X,Y)$ is
represented.}
\label{fig2}
\end{figure}

\begin{figure}
\onefigure[width=0.49\textwidth]{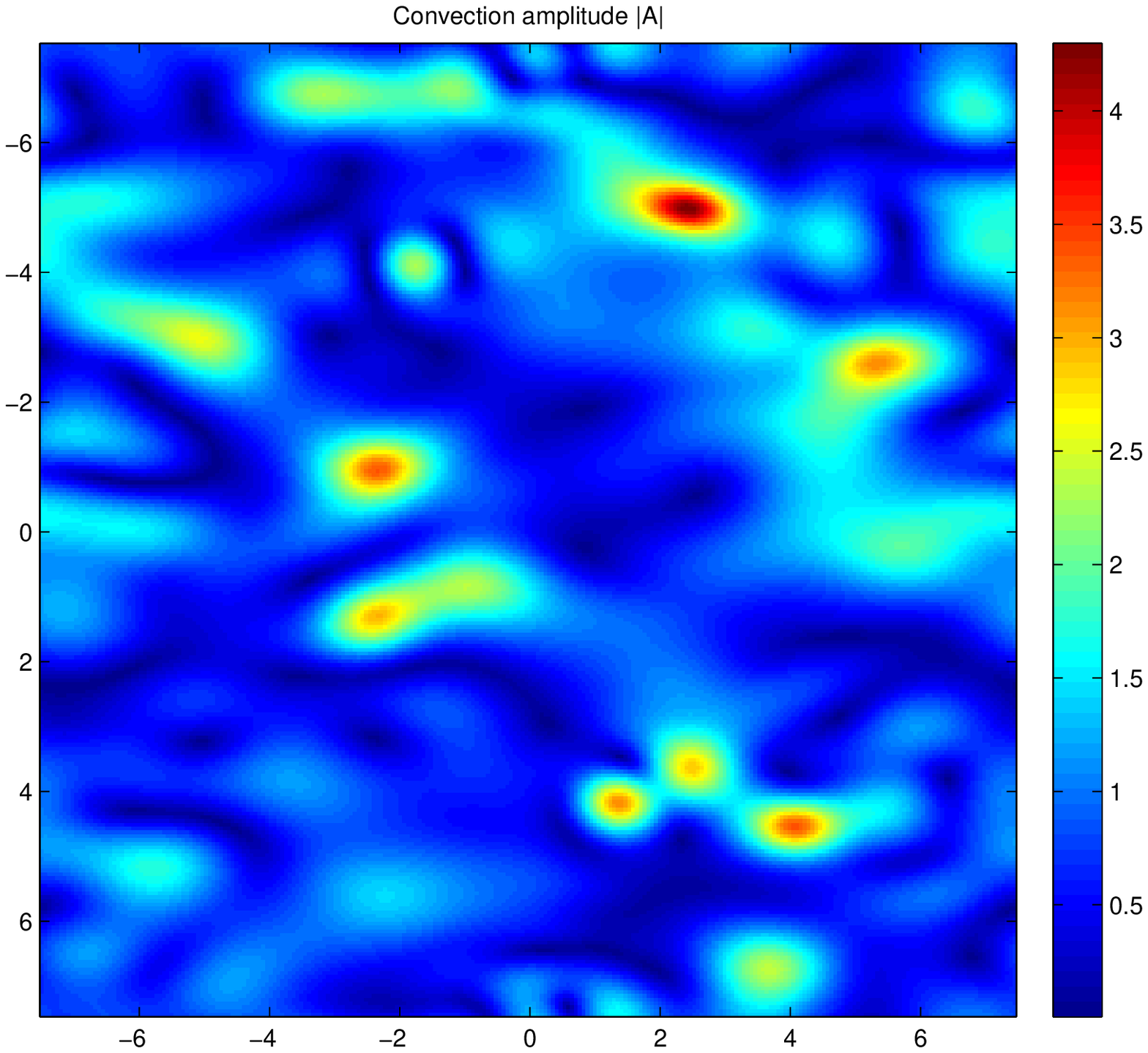}
\caption{Numerical solution of the one-mode equations in the
area $15\times 15$ for $T=30$ at $\omega=150000$.
On the figure the modulus of the amplitude $|A(T,X,Y)|$ is
represented. Initial conditions are an arbitrary noise of
the amplitude $10^{-4}$.}
\label{fig3}
\end{figure}

\begin{figure}
\onefigure[width=0.49\textwidth]{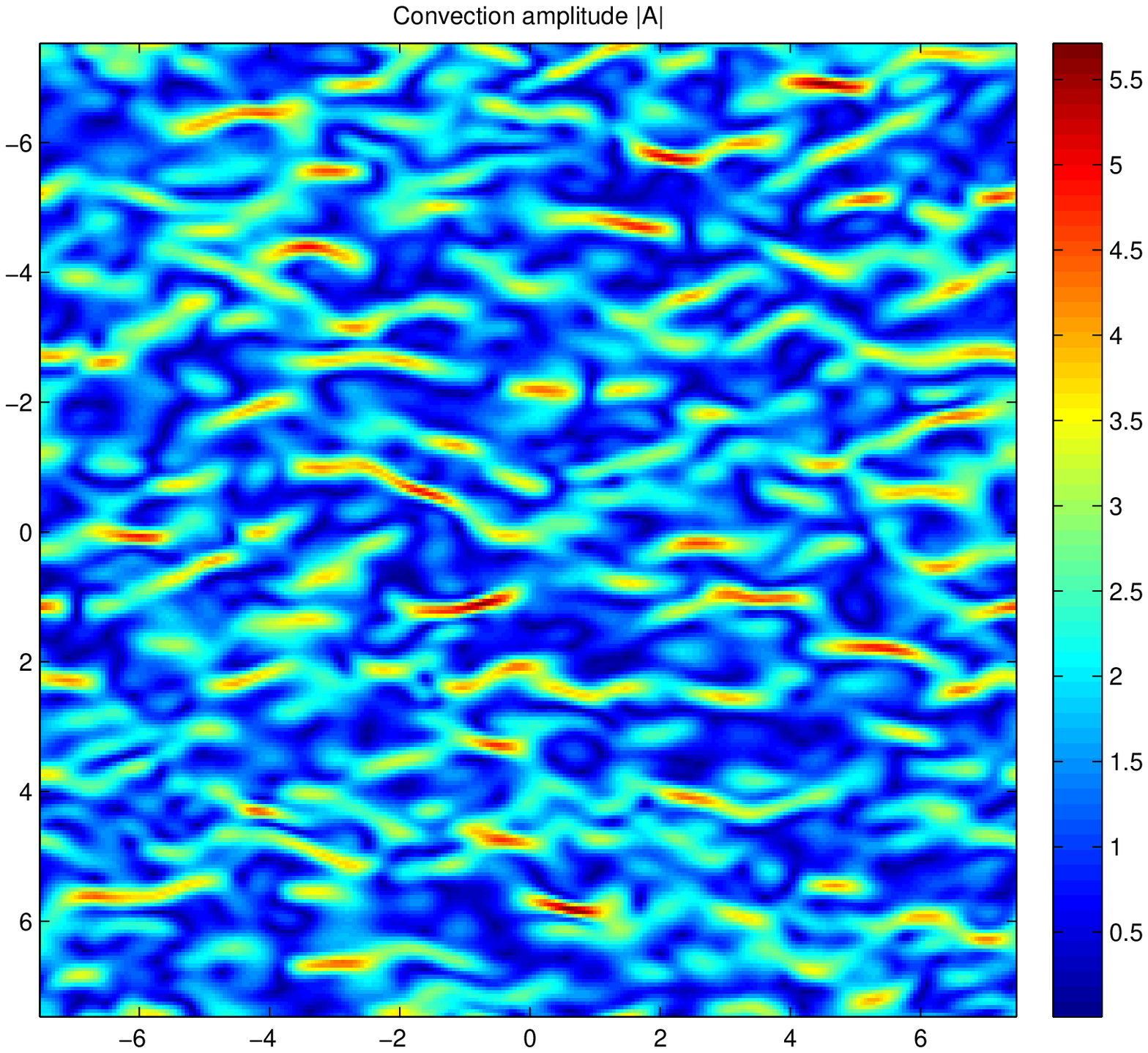}
\caption{Numerical solution of the one-mode equations in the
area $15\times 15$ for $T=30$ at $\omega=2000$.
On the figure the modulus of the amplitude $|A(T,X,Y)|$ is
represented. Initial conditions are an arbitrary noise of
the amplitude $10^{-4}$.}
\label{fig4}
\end{figure}

\begin{figure}
\onefigure[width=0.49\textwidth]{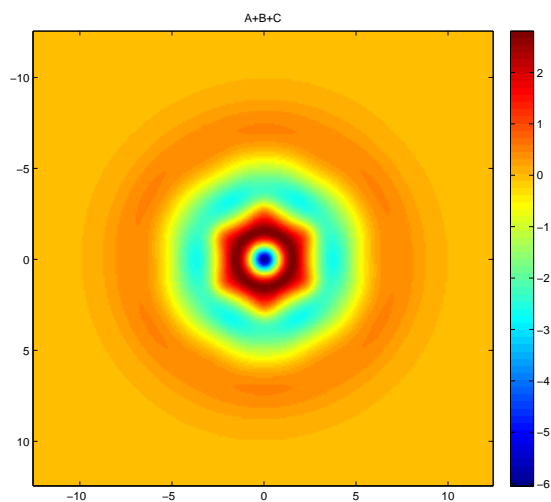}
\caption{Numerical solution of the tree-mode equations in the
area $25\times 25$ for $T=2.50$ at $\omega=20000$.
On the figure the sum of the amplitudes $Re(A(T,X,Y)+B(T,X,Y)+C(T,X,Y))$ is
represented. Initial conditions are $A=B=C=2\exp(-2(X^2+Y^2)).$}
\label{fig5}
\end{figure}

\begin{figure}
\onefigure[width=0.49\textwidth]{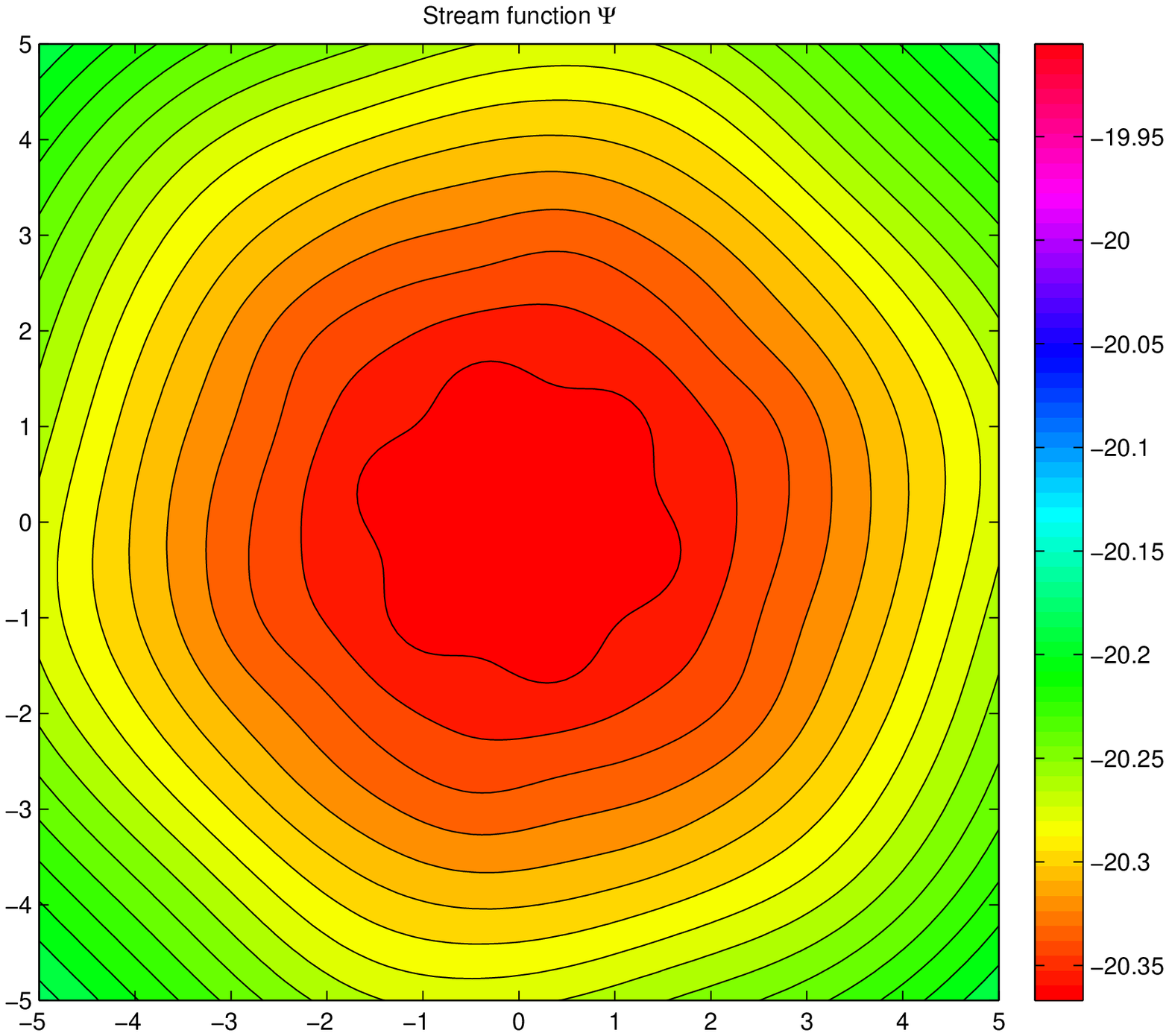}
\caption{Numerical solution of the tree-mode equations in the
area $25\times 25$ for $T=2.50$ at $\omega=20000$.
On the figure the stream function sum $\Psi(T,X,Y)$ is
represented. Initial conditions are $A=B=C=2\exp(-2(X^2+Y^2)).$}
\label{fig6}
\end{figure}

\subsection{The results of numerical simulation}

Numerical simulation for the Case 2 ($\omega = 20000$) shows, that
convection evolves from the initial arbitrary noise to the some
developed structure in a time of about $T = 15$. This structure has
a form of elongated ``clouds'', or ``sticks'' and ``spots'' for the
convective amplitude (see Fig.~\ref{fig1}) and a form of ``clouds''
for the respective stream function (see Fig.~\ref{fig2}).

Numerical simulation for the Case 3 ($\omega = 150000$) shows, that
convective patterns become more spot-like (see Fig.~\ref{fig3}) for
the noise initial conditions. And for the Case 1 ($\omega = 2000$)
the patterns have a form of threads or filaments (see
Fig.~\ref{fig4}).

All obtained patterns slowly evolve with time, and system never
reaches any stationary state. This is also true for the initial
conditions in the form of bell-like function. It was noticed that in
this case the system rather quickly (in a time of T = 15--35)
develops the condition of diffusion chaos, when the initial state is
destroyed and symmetrical convection becomes irregular in both space
and time. In this regime in some areas for a certain parameters peak
bursts of vorticity are noticed.

In the case of two or more modes the total spatial pattern of
convection appears as irregular alternation of convective cells of
various shapes. Wherein each mode and stream function of the
solution are qualitatively similar to the case of roll convection
(Fig. \ref{fig1}-\ref{fig2}). Of course, the solutions for each mode
nonlinearly interact with each other. As a result a ``curly''
structure arises, composed of curling threads pieces. For the
regular bell-like initial conditions mode interaction at short times
gives beautiful regular patterns. For the three-mode equations these
patterns may resemble for convection amplitude (see Fig.~\ref{fig5})
and for stream function (see Fig.~\ref{fig6}) famous Saturn's polar
hexagon.

There were cases when at some values of parameters in multi-mode
convection amplitude of the cells grew up to the formation of a
singular solution. For the regularization of such situations it was
sufficient to put into the equations minor amendments in the form of
terms of the fifth order in amplitude. A more detailed descriptions
and analysis of the results of numerical modeling of the equations
(\ref{eq_fin}) are beyond the scope of this article.

\section{An approach to the Saturns's polar hexagon simulation}

Saturn's hexagon is a persisting hexagonal cloud pattern around the
north pole of Saturn. The sides of the hexagon are about 13,800 km
long. The hexagon does not shift in longitude like other clouds in
the visible atmosphere. Saturn's polar hexagon discovery was made by
the Voyager mission in 1981--82, and it was revisited since 2006 by
the Cassini mission.

It is believed that the hexagon is described by some kind of
solitonic solution. Also it is stated that the hexagon forms where
there is a steep latitudinal gradient in the speed of the
atmospheric winds in Saturn's atmosphere. And the speed differential
and viscosity parameters should be within certain margins. If this
is not fulfilled the polygons don't arise, as at other likely
places, such as Saturn's South pole or the poles of Jupiter.

Obviously double-diffusive convection plays in the atmospheres of
such planets as Saturn or Jupiter an essential role. Here atmosphere
is a mixture of hydrogen with helium, and in the upper atmosphere
there exist a vertical negative gradient of temperature due to hot
lower layers. Thus we have a diffusive type of double-diffusive
convection in a rotation system. As a rule, rotation acts as one
more diffusive component, which gives actually a case of
triple-diffusive convection and complicates the analysis.
Nevertheless preliminary considerations show that at large Rayleigh
numbers (as in the case of Hexagon) such system behaves
qualitatively as the explored double-diffusive system near the Hopf
bifurcation points. So one can expect similar amplitude equations
for a slow variations of convective amplitude, but with the
different coefficients of such equations.

An exact derivation of amplitude equation for the Hexagon's case is
rather cumbersome task, but the obtained in this article results
allow to make some hints on possible steps in solving a task of
construction the equations having Hexagon as a solution.
As we noted, in the case of Hexagon one can expect amplitude
equations similar to (\ref{eq_fin}), but with additional terms. As
one can see, the solution (Fig.~\ref{fig5}-\ref{fig6}) for
three-mode equations qualitatively resembles Hexagon, but only on
not very large times. On large time the solution spreads over all
area, and its shape becomes more whimsical. So one should insert
into equations the stabilizing terms possibly taking into account
centrifugal forces and the curvature of the surface. Thus solution
will stay in the restricted area and have stable hexagonal form.
This shape itself is defined by tree interacting modes, and
coefficients of the equation should answer the question why
three-mode solution dominates over other multi-mode solutions.

\section{Conclusion}

The family of amplitude equations (\ref{eq_fin}) describing three-dimensional
double-diffusive convection in an infinite layer of fluid,
interacting with horizontal vorticity field is derived. The shape of
the convective cells is defined by a finite superposition of
roll-type modes.

For numerical simulation of the obtained systems of amplitude
equations we developed a few numerical schemes based on modern ETD
(exponential time differencing) pseudospectral methods~\cite{Cox}.
The software packages were written for simulation of roll-type convection and
convection with square and hexagonal type cells.

Numerical simulation has showed that the convection in the system
takes the form of convective ``spots'', ``sticks'' or ``filaments''
and elongated ``clouds'' as for the respective stream functions. In
the system quite rapidly (at time T = 15-35) a state of diffusive
chaos is developed, where the initial symmetric state is destroyed
and convection becomes irregular both in space and time. At the same
time in some areas there are bursts of vorticity.

The obtained results induce a deeper understanding of heat and mass
transfer processes in the ocean and the atmosphere, significantly
affecting the environment and migration of various impurities. These
results will help to describe more adequately the convective and
vortex structures that arise in physical systems with convective
instability, and may also be the basis for the construction of more
advanced models of systems with multi-component convection.

%----------------------------------------------


\begin{thebibliography}{20}

\bibitem{Getling}
\Name{Getling~A.~V.}
 \Book{Rayleigh-Benard Convection: Structures and Dynamics, Advanced Series in Nonlinear Dynamics}
 \Vol{11}
 \Publ{World Scientific, Singapore-River Edge, New Jersey}
 \Year{1998}
 \Page{245}.

\bibitem{Huppert}
\Name{Huppert~H.~E. \and Turner~J.~S.}
 \REVIEW{J. Fluid Mech.}{106}{1981}{299}

\bibitem{Radko}
\Name{Radko~T.}
 \Book{Double-diffusive convection}
 \Publ{Cambridge University Press}
 \Year{2013}
 \Page{344}.

\bibitem{KABC}
\Name{Kozitskiy~S.~B.}
 \Review{Phys. Rev. E}
 \Vol{72}
 \Year{2005}
 \Page{056309-1}

%\bibitem{Stommel}
%\Name{Stommel~H., Arons~A.~B. \and Blanchard D.}
% \REVIEW{Deep-Sea Res.}{3}{1956}{152}

%\bibitem{Knobloch}
%\Name{Knobloch~E., Moore~D.~R., Toomre~J. \and Weiss~N.~O.}
% \REVIEW{J. Fluid Mech.}{166}{1986}{409}

%\bibitem{Blue}
%\Name{Meca~E., Mercader~I., Batiste~O. \and Rami'rez-Piscina~L.}
% \REVIEW{Phys. Rev. Lett.}{92}{2004}{234501-1}

\bibitem{Newell}
\Name{Newell~A.~C. \and Whitehead~J.~A.}
 \REVIEW{J. Fluid Mech.}{38}{1968}{279}

\bibitem{Bretherton}
\Name{Bretherton~C.~S. \and Spiegel~E.~A.}
\REVIEW{Phys. Lett.}{96A}{1983}{152}

\bibitem{Zippelius}
\Name{Zippelius~A. \and Siggia~E.~D.} \REVIEW{Phys.
Fluids}{26}{1983}{2905}

\bibitem{K10}
\Name{Kozitskiy~S.~B.}
\REVIEW{J. Appl. Mech. and Tech. Phys.}{41(3)}{2000}{429}

\bibitem{K11}
\Name{Kozitskiy~S.~B.}
\REVIEW{Vestn. Udmurt. Univ. Mat. Mekh. Komp'yut. Nauki}{3}{2008}{46}

\bibitem{K15}
\Name{Kozitskiy~S.~B.}
\REVIEW{Vestn. Udmurt. Univ. Mat. Mekh. Komp'yut. Nauki}{4}{2010}{13}

\bibitem{K16}
\Name{Kozitskiy~S.~B.}
\REVIEW{Vestn. Udmurt. Univ. Mat. Mekh. Komp'yut. Nauki}{4}{2012}{23}

\bibitem{Lan}
\Name{Landau~L.~D. \and Lifshits~E.~M.}
 \Book{Fluid Mechanics, Course of Theoretical Physics Vol. 6.}
 \Vol{6}
 \Publ{Pergamon Press, Oxford}
 \Year{1999}
 \Page{539}.

\bibitem{Weiss}
\Name{Weiss~N.~O.}
\REVIEW{J. Fluid Mech.}{108}{1981}{247}

\bibitem{BB}
\Name{Balmforth~N.~J. \and Biello~J.~A.} \REVIEW{J. Fluid
Mech.}{375}{1998}{203}


\bibitem{Nayfeh}
\Name{Nayfeh~A.~H.}
 \Book{Introduction to perturbation techniques.}
 \Publ{John Wiley \& Sons, New York-Chichester-Brisbane-Toronto}
 \Year{1993}
 \Page{536}.

\bibitem{Dry}
\Name{Cooke~K.~L.}
\REVIEW{J. Math. Anal. and Appl.}{24}{1968}{372}

\bibitem{Cox}
\Name{Cox~S.~M. \and Matthews~P.~C.}
\REVIEW{J. Comput. Phys.}{176}{2002}{430}

\end{thebibliography}
\end{document}